\documentclass[prb,showpacs,twocolumn]{revtex4}%
\usepackage{graphicx}
\usepackage{dcolumn}
\usepackage{bm}
\usepackage{amsmath}
\usepackage{amsfonts}
\usepackage{amssymb}%
\providecommand{\U}[1]{\protect\rule{.1in}{.1in}}
\begin{document}
\title{$\emph{GW}$\textbf{ approximation with self-screening correction}}
\author{F. Aryasetiawan$^{1,2}$, R. Sakuma$^{1,2}$, and K. Karlsson$^{3}$}
\affiliation{$^{1}$Graduate School of Advanced Integration Science, Chiba University, 1-33
Yayoi-cho, Inage-ku, Chiba-shi, Chiba, 263-8522 Japan,}
\affiliation{$^{2}$Japan Science and Technology Agency, CREST, Kawaguchi, Saitama 332-0012, Japan}
\affiliation{$^{3}$Department of Life Sciences,}
\affiliation{H\"{o}gskolan i Sk\"{o}vde, 54128 Sk\"{o}vde, Sweden}

\begin{abstract}
The \emph{GW} approximation takes into account electrostatic self-interaction
contained in the Hartree potential through the exchange potential. However, it
has been known for a long time that the approximation contains self-screening
error as evident in the case of the hydrogen atom. When applied to the
hydrogen atom, the \emph{GW} approximation does not yield the exact result for
the electron removal spectra because of the presence of self-screening: the
hole left behind is erroneously screened by the only electron in the system
which is no longer present. We present a scheme to take into account
self-screening and show that the removal of self-screening is equivalent to
including exchange diagrams, as far as self-screening is concerned. The scheme
is tested on a model hydrogen dimer and it is shown that the scheme yields the
exact result to second order in $(U_{0}-U_{1})/2t$ where $U_{0}$ and $U_{1}$
are respectively the onsite and offsite Hubbard interaction parameters and $t$
the hopping parameter.

\end{abstract}

\maketitle

\section{Introduction}

In the Hartree approximation \cite{Hartree}, a system of electrons move in a
common potential arising from the electrostatic field of the electrons, in
addition to the external field. In this approximation, a given electron
experiences the electrostatic potential from the other electrons as well as
from itself because the common potential or the Hartree potential contains the
field from the electron itself. This unphysical self-interaction is removed
when exchange interaction is included, leading to the Hartree-Fock
approximation (HFA) \cite{Fock}. In density functional theory \cite{Kohn},
Perdew and Zunger introduced the concept of self-interaction correction
\cite{Perdew1981} to remove a similar problem in the local density
approximation (LDA).

For many-electron systems, such as solids, it is well known that the HFA is
not satisfactory because it completely neglects screening which is very
crucial in describing the electronic structure of many-electron systems. Thus,
for example, the Hartree-Fock band gaps of semiconductors and insulators are
much too wide and when the HFA is applied to metals the density of states at
the Fermi level becomes unphysically zero due to the logarithmic singularity
in the derivative of the one-particle energy with respect to the $k$-vector at
$k=k_{F}$ \cite{Ashcroft}. The simplest known and successful method beyond the
HFA that cures the band-gap problem and the anomaly of the HFA in metals is
the $GW$ approximation (GWA) \cite{Hedin1965,ferdi1998}. The GWA includes the
effects of frequency-dependent screening from first principles and the
self-energy in space-time representation is approximated by a product of the
Green function $G$ and the screened interaction $W.$

The GWA includes the exchange potential so that it is self-interaction free.
However, it is contaminated by "self-screening", namely, an electron screens
itself, analogous to "self-interaction" where an electron interacts with its
own electrostatic field. This undesirable self-screening effect has been a
long-standing problem and thought to be a source of significant errors in the
electronic structure. The self-screening problem may be illustrated by the
famous case of the hydrogen atom. Since there is only one electron, it is
clear that the one-particle removal energy or the hole energy is simply given
by 13.6 eV, the $1s$ orbital energy. The Hartree approximation applied to the
hydrogen atom would yield a too low removal energy due to the self-interaction
error while the HFA would give the correct result. Embarassingly, when the GWA
is applied to the hydrogen atom, it yields a wrong result because, as a
consequence of self-screening, the correlation part of the \emph{hole}
self-energy in the GWA is not zero \cite{Godby2007}. Evidently, since there is
only one electron, upon removal of the electron there are no other electrons
that can screen the remaining hole so that the hole self-energy ought to be zero.

The self-screening error is believed to be responsible for a number of
well-known problems. It has been suspected for a long time that the presence
of self-screening in the $GW$ self-energy may be responsible for errors in the
quasiparticle energies of localized states. It has been found that $GW$
quasiparticle energies of core or semicore states usually lie above the
experimental values. It is argued that in the HFA the quasiparticle energies
are too low due to the absence of screening and when screening is taken into
account within the GWA, these energies are pushed up too high, an indication
of overscreening due to self-screening. In molecules, a recent comprehensive
and systematic study of 34 molecules has found that the GWA overscreens the
Hartree-Fock ionization potential leading to underestimation by $0.4\sim0.5$
eV compared to experiment \cite{Rostgaard2010}. In many materials, the energy
position of the core or semicore states is usually too high in the LDA due to
self-interaction. $GW$ calculations on the 3d semicore states of a number of
semiconductors such as GaAs and ZnSe improve the LDA results but the remaining
error is still significant \cite{ferdi1996}. It is very likely that this error
owes its origin from self-screening. From physical consideration the
self-screening error is expected to be significant when the states are rather
localized but less important in extended states

In this paper, we develop a new scheme which aims at correcting the
self-screening error in the $GW$ self-energy as well as the linear
density-density response function within the random-phase approximation (RPA)
\cite{Pines}. An interesting consequence of the proposed scheme is the fact
that the screened interaction $W$ becomes explicitly spin dependent, in
contrast to the original GWA where the screened interaction is spin
independent. We also furnish a theoretical support for the scheme by showing
from diagrammatic consideration that the removal of the self-screening terms
is partially equivalent to adding exchange diagrams. In other words, the
self-screening terms are cancelled by corresponding terms in the exchange diagrams.

As an illustration of our scheme, we calculate the bonding-antibonding gap of
a model hydrogen dimer. We have chosen this model because the exact result is
known allowing for rigorous comparison. Moreover, the calculations can be
performed analytically so that possible numerical errors are eliminated and
the simplicity of the system permits us to analyze the results without
unnecessary complicating factors. It is found that the self-screening
corrected GWA reproduces the exact result to order $[(U_{0}-U_{1})/2t]^{2}$,
where $U_{0}$ and $U_{1}$ are respectively the onsite and offsite Coulomb
energies and $t$ is the hopping integral.

\section{$GW$ approximation with self-screening correction}

\subsection{Theory}

The first step of the procedure is to decompose the non-interacting Green
function into its orbital components:
\begin{equation}
G_{\sigma}^{0}(\mathbf{r,r}^{\prime};\omega)=\sum_{n}g_{n\sigma}%
(\mathbf{r,r}^{\prime};\omega), \label{G0}%
\end{equation}%
\begin{equation}
g_{n\sigma}(\mathbf{r,r}^{\prime};\omega)=\frac{\varphi_{n\sigma}%
(\mathbf{r})\varphi_{n\sigma}^{\ast}(\mathbf{r}^{\prime})}{\omega
-\varepsilon_{n\sigma}}, \label{gn}%
\end{equation}

\begin{equation}
h\varphi_{n\sigma}=\varepsilon_{n\sigma}\varphi_{n\sigma}, \label{hphi}%
\end{equation}
where $h$ is a one-particle Hamiltonian, $\varepsilon_{n\sigma}\rightarrow
\varepsilon_{n\sigma}+i\delta$ for an occupied state and $\varepsilon
_{n\sigma}\rightarrow\varepsilon_{n\sigma}-i\delta$ for an unoccupied state.
We refer to $\left\{  g_{n\sigma}\right\}  $ as orbital Green functions. In
the GWA the self-energy is given by, using a non-interacting $G^{0}$,%

\begin{align}
\Sigma_{\sigma}(\mathbf{r}t,\mathbf{r}^{\prime}t^{\prime})  &  =iG_{\sigma
}^{0}(\mathbf{r}t,\mathbf{r}^{\prime}t^{\prime})W(\mathbf{r}^{\prime}%
t^{\prime},\mathbf{r}t)\nonumber\\
&  =i\sum_{m}g_{m\sigma}(\mathbf{r}t,\mathbf{r}^{\prime}t^{\prime
})W(\mathbf{r}^{\prime}t^{\prime},\mathbf{r}t), \label{Sigma}%
\end{align}
where $W$ is the screened interaction%

\begin{equation}
W=\epsilon^{-1}v \label{W}%
\end{equation}
with $\epsilon$ being the dielectric matrix.

\begin{figure}
[ptb]
\begin{center}
\includegraphics[
height=1.7443in,
width=2.8435in
]%
{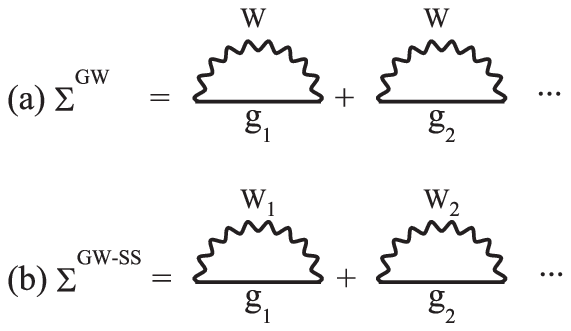}%
\caption{Comparison between the self-energy diagrams in the conventional GWA
(a) and the GWA with self-screening correction (b). In the latter, the
screened interaction depends both on the orbital and spin of the electron
represented by $g_{n}$, as discussed in the text. }%
\label{diagrams}%
\end{center}
\end{figure}

Consider an electron occupying an orbital $\varphi_{m\sigma}$ propagating from
$(\mathbf{r}^{\prime}t^{\prime})$ to $(\mathbf{r}t)$ represented by
$g_{m\sigma}$. Another electron with the \emph{same spin} cannot occupy the
orbital $\varphi_{m\sigma}$ and therefore $g_{m\sigma}$ should not participate
in the screening process during the propagation of the electron. Therefore the
screened interaction $W$ should be calculated using a polarization propagator
that does not include $g_{m\sigma}$. However, an electron in the same orbital
but with opposite spin can naturally participate in the screening process. The
self-energy then takes the following form:%

\begin{equation}
\Sigma_{\sigma}(\mathbf{r}t,\mathbf{r}^{\prime}t^{\prime})=i\sum_{m}%
g_{m\sigma}(\mathbf{r}t,\mathbf{r}^{\prime}t^{\prime})W_{m\sigma}%
(\mathbf{r}^{\prime}t^{\prime},\mathbf{r}t), \label{Sigma_SSC}%
\end{equation}
where%

\begin{equation}
W_{m\sigma}=v+vR_{m\sigma}v=v+W_{m\sigma}^{c}, \label{Wm}%
\end{equation}

\begin{equation}
R_{m\sigma}=P_{m\sigma}+P_{m\sigma}vR_{m\sigma}. \label{Rm}%
\end{equation}
The polarization $P_{m\sigma}$ is defined as the polarization without
$g_{m\sigma}$, i.e., no Green function line in $P_{m\sigma}$ contains
$g_{m\sigma}$. In other words,%

\begin{equation}
P_{m\sigma}=-i(G_{m\sigma}G_{m\sigma}+G_{-\sigma}G_{-\sigma}), \label{Pm}%
\end{equation}
where $G_{m\sigma}$ is the Green function without $g_{m\sigma}$, namely,%
\begin{equation}
G_{m\sigma}=G_{\sigma}-g_{m\sigma}. \label{Gm}%
\end{equation}
In Fig. \ref{diagrams} the self-energy diagrams corresponding to
(\ref{Sigma_SSC}) are compared with the conventional $GW$ diagrams.%

The correlation part of the $GW$ self-energy with self-screening correction is
given by%

\begin{align}
&  \Sigma_{\sigma}^{c}(\mathbf{r,r}^{\prime};\omega)\nonumber\\
&  =i\sum_{m}\int\frac{d\omega^{\prime}}{2\pi}g_{m\sigma}(\mathbf{r,r}%
^{\prime};\omega+\omega^{\prime})W_{m}^{c}(\mathbf{r}^{\prime}\mathbf{,r}%
;\omega^{\prime})\nonumber\\
&  =i\sum_{m}\int\frac{d\omega^{\prime}}{2\pi}\frac{\varphi_{m\sigma
}(\mathbf{r})\varphi_{m\sigma}^{\ast}(\mathbf{r}^{\prime})W_{m}^{c}%
(\mathbf{r}^{\prime}\mathbf{,r};\omega^{\prime})}{\omega+\omega^{\prime
}-\varepsilon_{m\sigma}+i\delta\text{sgn}(\varepsilon_{m\sigma}-\mu)}.
\label{Sigmac}%
\end{align}
Writing the correlation part of the screened interaction, $W^{c}$, in its
spectral representation%

\begin{align}
W_{m\sigma}^{c}(\mathbf{r}^{\prime}\mathbf{,r};\omega^{\prime})  &
=\int_{-\infty}^{0}d\omega^{\prime\prime}\frac{D_{m\sigma}(\mathbf{r}^{\prime
}\mathbf{,r};\omega^{\prime\prime})}{\omega^{\prime}-\omega^{\prime\prime
}-i\delta}\nonumber\\
&  +\int_{0}^{\infty}d\omega^{\prime\prime}\frac{D_{m\sigma}(\mathbf{r}%
^{\prime}\mathbf{,r};\omega^{\prime\prime})}{\omega^{\prime}-\omega
^{\prime\prime}+i\delta}, \label{Wc}%
\end{align}
the frequency integral over $\omega^{\prime}$ can be performed analytically.
The correlation part of the self-energy may be divided into two parts,
$\Sigma^{\text{occ}}$ and $\Sigma^{\text{unocc}}$:%

\begin{align}
&  \Sigma_{\sigma}^{\text{occ}}(\mathbf{r,r}^{\prime};\omega)\nonumber\\
&  =\sum_{m}^{\text{occ}}\int_{0}^{\infty}d\omega^{\prime\prime}\frac
{\varphi_{m\sigma}(\mathbf{r})D_{m\sigma}(\mathbf{r}^{\prime}\mathbf{,r}%
;\omega^{\prime\prime})\varphi_{m\sigma}^{\ast}(\mathbf{r}^{\prime})}%
{\omega+\omega^{\prime\prime}-\varepsilon_{m\sigma}-i\delta},
\label{Sigma_occ}%
\end{align}

\begin{align}
&  \Sigma_{\sigma}^{\text{unocc}}(\mathbf{r,r}^{\prime};\omega)\nonumber\\
&  =\sum_{m}^{\text{unocc}}\int_{0}^{\infty}d\omega^{\prime\prime}%
\frac{\varphi_{m\sigma}(\mathbf{r})D_{m\sigma}(\mathbf{r}^{\prime}%
\mathbf{,r};\omega^{\prime\prime})\varphi_{m\sigma}^{\ast}(\mathbf{r}^{\prime
})}{\omega-\omega^{\prime\prime}-\varepsilon_{m\sigma}+i\delta},
\label{Sigma_unocc}%
\end{align}
$D_{m\sigma}$ is the spectral function of $W_{m\sigma}^{c}$ and we have used
the relations%

\begin{equation}
D_{m\sigma}(-\omega)=-D_{m\sigma}(\omega),\ W_{m\sigma}^{c}(-\omega
)=W_{m\sigma}^{c}(\omega) \label{symmetry}%
\end{equation}
and%

\begin{equation}
D_{m\sigma}(\omega)=-\frac{1}{\pi}\operatorname{Im}W_{m\sigma}^{c}%
(\omega)\text{sgn}(\omega). \label{D}%
\end{equation}
The corresponding expressions for the self-energy in the conventional GWA are
the same as above except that $D_{m\sigma}$ is replaced by the spectral
function of $W$ instead.

It is worth noting that the self-screening correction introduces spin
dependence in the screened interaction $W$ as can be seen in (\ref{Sigma_SSC}%
). Each electron experiences a different screened interaction $W_{m\sigma}$
which is not only orbital dependent but also spin dependent according to the
orbital occupied by the electron as well as the spin of the electron.

Since in the exact set of Hedin's equations the screened interaction $W$ is
spin independent the appearance of a spin-dependent screened interaction seems
unnecessary. It is interesting to make comparison with density functional
theory. In principle, the total energy is obtainable from the ground-state
electron density, which is the sum of spin up and down components. In
practice, for spin-polarized systems it is more favorable to introduce the
spin variable and regard the total energy as a functional of the up and down
spin densities. The separation of the density into the up and down components
mimics the true system and captures the essential physics so that a relatively
simple approximation, such as the local spin density approximation, still
works well. A presumably much more complicated functional would be required to
achieve the same level of accuracy for the total energy if the total density
were to be used instead. A similar situation arise in our case, where the
orbital and spin-dependent screened interactions closely mirror the physical
situation and thereby promotes a better self-energy within the simple GWA. If
we kept the conventional screened interaction, we would need to include
exchange diagrams as vertex corrections to cancel the self-screening terms, as
shown in a later section. It is much simpler to remove the self-screening
terms than to include vertex corrections.

\subsection{Self-screening correction in extended states}

For extended states, the self-screening correction tends to vanish. However,
from the physical point of view, we expect that the self-screening correction
is significant when the state originates from a localized orbital such as the
case with the states originating from the 3d or 4f orbitals. Consider
expanding a given Bloch state in its Wannier representation \cite{Marzari1997}%

\begin{equation}
\psi_{\mathbf{k}n\sigma}(\mathbf{r})=\frac{1}{\sqrt{N}}\sum_{\mathbf{R}}%
\exp(i\mathbf{k\cdot R})\chi_{\mathbf{R}n\sigma}(\mathbf{r}). \label{Bloch}%
\end{equation}
The Green function is, with $\varepsilon_{\mathbf{k}n\sigma}\rightarrow
\varepsilon_{\mathbf{k}n\sigma}+i\eta$ for occupied states and $\varepsilon
_{\mathbf{k}n\sigma}\rightarrow\varepsilon_{\mathbf{k}n\sigma}-i\eta$ for
unoccupied states,%

\begin{align}
&  G_{\sigma}^{0}(\mathbf{r,r}^{\prime};\omega)\nonumber\\
&  =\sum_{kn}\frac{\psi_{\mathbf{k}n\sigma}(\mathbf{r})\psi_{\mathbf{k}%
n\sigma}^{\ast}(\mathbf{r}^{\prime})}{\omega-\varepsilon_{\mathbf{k}n\sigma}%
}\nonumber\\
&  =\frac{1}{N}\sum_{\mathbf{k}n}\sum_{\mathbf{RR}^{\prime}}\frac
{\exp[i\mathbf{k\cdot(R-R}^{\prime})]\chi_{\mathbf{R}n\sigma}(\mathbf{r}%
)\chi_{\mathbf{R}^{\prime}n\sigma}^{\ast}(\mathbf{r}^{\prime})}{\omega
-\varepsilon_{\mathbf{k}n\sigma}}\nonumber\\
&  =\frac{1}{N}\sum_{\mathbf{k}n}\sum_{\mathbf{R\neq R}^{\prime}}\frac
{\exp[i\mathbf{k}\cdot(\mathbf{R-R}^{\prime})]\chi_{\mathbf{R}n\sigma
}(\mathbf{r})\chi_{\mathbf{R}^{\prime}n\sigma}^{\ast}(\mathbf{r}^{\prime}%
)}{\omega-\varepsilon_{\mathbf{k}n\sigma}}\nonumber\\
&  +\frac{1}{N}\sum_{\mathbf{k}n}\sum_{\mathbf{R}}\frac{\chi_{\mathbf{R}%
n\sigma}(\mathbf{r})\chi_{\mathbf{R}n\sigma}^{\ast}(\mathbf{r}^{\prime}%
)}{\omega-\varepsilon_{\mathbf{k}n\sigma}}. \label{GWannier}%
\end{align}
We apply the self-screening correction to the component of $G_{\sigma}^{0}$
corresponding to $\mathbf{R=R}^{\prime}$. As before we define%

\begin{equation}
G_{n\sigma}^{0}=G_{\sigma}^{0}-g_{n\sigma}, \label{G0n}%
\end{equation}

\begin{align}
g_{n\sigma}(\mathbf{r,r}^{\prime};\omega)  &  =\frac{1}{N}\sum_{\mathbf{k}%
}\sum_{\mathbf{R}}\frac{\chi_{\mathbf{R}n\sigma}(\mathbf{r})\chi
_{\mathbf{R}n\sigma}^{\ast}(\mathbf{r}^{\prime})}{\omega-\varepsilon
_{\mathbf{k}n\sigma}}\nonumber\\
&  =\frac{1}{N}\sum_{\mathbf{R}}\chi_{\mathbf{R}n\sigma}(\mathbf{r}%
)g_{n\sigma}(\omega)\chi_{\mathbf{R}n\sigma}^{\ast}(\mathbf{r}^{\prime}),
\label{gnWannier}%
\end{align}
where%

\begin{equation}
g_{n\sigma}(\omega)=\sum_{\mathbf{k}}\frac{1}{\omega-\varepsilon
_{\mathbf{k}n\sigma}}. \label{gnw}%
\end{equation}
In practice, there may be a problem due to the non-analytic behaviour of
$\varepsilon_{\mathbf{k}n\sigma}$ as a function of $\mathbf{k}$. For very
narrow band such as the one formed by semicore states we may make the
following approximation%

\begin{equation}
g_{n\sigma}(\mathbf{r,r}^{\prime};\omega)\approx\frac{1}{N}\sum_{\mathbf{R}%
}\frac{\chi_{\mathbf{R}n\sigma}(\mathbf{r})\chi_{\mathbf{R}n\sigma}^{\ast
}(\mathbf{r}^{\prime})}{\omega-\left\langle \varepsilon_{n\sigma}\right\rangle
} \label{gnapprox}%
\end{equation}

\begin{equation}
\left\langle \varepsilon_{n\sigma}\right\rangle =\sum_{\mathbf{k}}%
\varepsilon_{\mathbf{k}n\sigma}. \label{eav}%
\end{equation}
For a given site $R$ the orbital Green function $g_{n\sigma}$ is confined to
the site and it is equivalent to a core state Green function.

\subsection{Theoretical justification of self-screening correction}

\begin{figure}
[ptb]
\begin{center}
\includegraphics[
height=1.2419in,
width=2.0176in
]%
{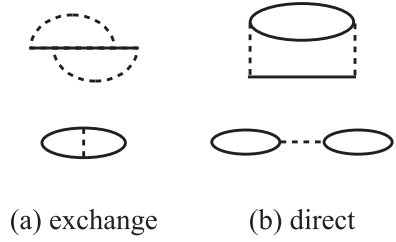}%
\caption{The second-order exhange and direct self-energy and polarization
diagrams. As shown in the text, the exchange diagrams cancel the
self-screening terms in the direct diagrams.}%
\label{xd}%
\end{center}
\end{figure}

Here we show that removing the self-screening terms in the self-energy is
partially equivalent to adding vertex corrections in the form of exchange
diagrams. It can be shown that the self-screening terms are cancelled by the
corresponding terms in the exchange diagrams in a similar fashion as for the
first-order self-energy or the HFA. We will illustrate the idea for the
second-order self-energy but it is clear that the argument applies to any
order. The second-order exchange and direct diagrams are shown in the upper
part of Fig. \ref{xd}.%

According to the Feynman rules \cite{Fetter}, the second-order exchange
self-energy for a given spin is%

\begin{align}
\Sigma^{x}(x_{1},x_{2})  &  =(i)^{2}\int dx_{3}dx_{4}G(x_{1},x_{3}%
)G(x_{3},x_{4})\nonumber\\
&  \times G(x_{4},x_{2})v(x_{1}-x_{4})v(x_{3}-x_{2}), \label{Sigma_x}%
\end{align}
where $x=(\mathbf{r},t)$ and%

\begin{equation}
v(x-x^{\prime})=v(\mathbf{r-r}^{\prime})\delta(t-t^{\prime}). \label{vcoul}%
\end{equation}
Fourier transformation with respect to $\tau=t_{1}-t_{2}$ yields%

\begin{align}
\Sigma^{x}(\mathbf{r}_{1}\mathbf{,r}_{2};\omega)  &  =-\int d^{3}r_{3}%
d^{3}r_{4}\int\frac{d\omega_{1}d\omega_{2}}{(2\pi)^{2}}G(\mathbf{r}%
_{1},\mathbf{r}_{3};\omega_{1})\nonumber\\
&  \times G(\mathbf{r}_{3},\mathbf{r}_{4};\omega_{2})G(\mathbf{r}%
_{4},\mathbf{r}_{2};\omega-\omega_{1}+\omega_{2})\ \nonumber\\
&  \times v(\mathbf{r}_{1}-\mathbf{r}_{4})v(\mathbf{r}_{3}-\mathbf{r}_{2}).
\label{Sigma_xw}%
\end{align}
Using a non-interacting Green function of a given spin%

\begin{equation}
G^{0}(\mathbf{r,r}^{\prime};\omega)=\sum_{n}^{\text{occ}}\frac{\varphi
_{n}(\mathbf{r})\varphi_{n}^{\ast}(\mathbf{r}^{\prime})}{\omega-\varepsilon
_{n}-i\delta}+\sum_{m}^{\text{unocc}}\frac{\varphi_{m}(\mathbf{r})\varphi
_{m}^{\ast}(\mathbf{r}^{\prime})}{\omega-\varepsilon_{m}+i\delta}, \label{G00}%
\end{equation}
we can perform the frequency integral over $\omega_{2}$ using Cauchy's theorem
by closing the contour either in the upper or lower plane:%

\begin{align}
&  \int\frac{d\omega_{2}}{2\pi}G^{0}(\mathbf{r}_{3},\mathbf{r}_{4};\omega
_{2})G^{0}(\mathbf{r}_{4},\mathbf{r}_{2};\omega-\omega_{1}+\omega
_{2})\nonumber\\
&  =i\sum_{n}^{\text{occ}}\sum_{m}^{\text{unocc}}\left\{  \frac{\varphi
_{n}(\mathbf{r}_{3})\varphi_{n}^{\ast}(\mathbf{r}_{4})\varphi_{m}%
(\mathbf{r}_{4})\varphi_{m}^{\ast}(\mathbf{r}_{2})}{\omega-\omega
_{1}+\varepsilon_{n}-\varepsilon_{m}+i\delta}\right. \nonumber\\
&  \ \ \ \ \ \ \ \ \ \ \ \ \ \ \ \ \ \ \ \left.  +\frac{\varphi_{m}%
(\mathbf{r}_{3})\varphi_{m}^{\ast}(\mathbf{r}_{4})\varphi_{n}(\mathbf{r}%
_{4})\varphi_{n}^{\ast}(\mathbf{r}_{2})}{\omega-\omega_{1}+\varepsilon
_{m}-\varepsilon_{n}-i\delta}\right\}  . \label{int1}%
\end{align}
Similarly, integrating over $\omega_{1}$ we find%

\begin{align}
&  \Sigma^{x}(\mathbf{r}_{1},\mathbf{r}_{2};\omega)\nonumber\\
&  =-\sum_{n}^{\text{occ}}\sum_{m}^{\text{unocc}}\int d^{3}r_{3}d^{3}%
r_{4}\ v(\mathbf{r}_{1}-\mathbf{r}_{4})v(\mathbf{r}_{3}-\mathbf{r}%
_{2})\nonumber\\
&  \times\left\{  \sum_{k}^{\text{occ}}\frac{\varphi_{k}(\mathbf{r}%
_{1})\varphi_{k}^{\ast}(\mathbf{r}_{3})\varphi_{m}(\mathbf{r}_{3})\varphi
_{m}^{\ast}(\mathbf{r}_{4})\varphi_{n}(\mathbf{r}_{4})\varphi_{n}^{\ast
}(\mathbf{r}_{2})}{\omega-\varepsilon_{k}+\varepsilon_{m}-\varepsilon
_{n}-i\delta}\right. \nonumber\\
&  +\left.  \sum_{k}^{\text{unocc}}\frac{\varphi_{k}(\mathbf{r}_{1}%
)\varphi_{k}^{\ast}(\mathbf{r}_{3})\varphi_{n}(\mathbf{r}_{3})\varphi
_{n}^{\ast}(\mathbf{r}_{4})\varphi_{m}(\mathbf{r}_{4})\varphi_{m}^{\ast
}(\mathbf{r}_{2})}{\omega-\varepsilon_{k}+\varepsilon_{n}-\varepsilon
_{m}+i\delta}\right\}  . \label{Sx}%
\end{align}

The second-order direct self-energy is%

\begin{align}
\Sigma^{d}(x_{1},x_{2})  &  =-(i)^{2}\int dx_{3}dx_{4}G(x_{1},x_{2}%
)G(x_{3},x_{4})G(x_{4},x_{3})\nonumber\\
&  \times v(x_{1}-x_{4})v(x_{3}-x_{2}),
\end{align}
and we have considered the direct term with all $G$ having the same spin since
this is the term that contains self-screening. Its Fourier transform is given by%

\begin{align}
\Sigma^{d}(\mathbf{r}_{1},\mathbf{r}_{2};\omega)  &  =\int d^{3}r_{3}%
d^{3}r_{4}\int\frac{d\omega_{1}d\omega_{2}}{(2\pi)^{2}}\times G(\mathbf{r}%
_{1},\mathbf{r}_{2};\omega_{1})\nonumber\\
&  \times G(\mathbf{r}_{3},\mathbf{r}_{4};\omega_{2})G(\mathbf{r}%
_{4},\mathbf{r}_{3};\omega-\omega_{1}+\omega_{2})\nonumber\\
&  \times v(\mathbf{r}_{1}-\mathbf{r}_{4})v(\mathbf{r}_{3}-\mathbf{r}_{2}),
\end{align}
which can be calculated analytically as in the exchange case yielding%

\begin{align}
&  \Sigma^{d}(\mathbf{r}_{1},\mathbf{r}_{2};\omega)\nonumber\\
&  =\sum_{n}^{\text{occ}}\sum_{m}^{\text{unocc}}\int d^{3}r_{3}d^{3}%
r_{4}\ v(\mathbf{r}_{1}-\mathbf{r}_{4})v(\mathbf{r}_{3}-\mathbf{r}%
_{2})\nonumber\\
&  \times\left\{  \sum_{k}^{\text{occ}}\frac{\varphi_{k}(\mathbf{r}%
_{1})\varphi_{k}^{\ast}(\mathbf{r}_{2})\varphi_{m}(\mathbf{r}_{3})\varphi
_{m}^{\ast}(\mathbf{r}_{4})\varphi_{n}(\mathbf{r}_{4})\varphi_{n}^{\ast
}(\mathbf{r}_{3})}{\omega-\varepsilon_{k}+\varepsilon_{m}-\varepsilon
_{n}-i\delta}\right. \nonumber\\
&  +\left.  \sum_{k}^{\text{unocc}}\frac{\varphi_{k}(\mathbf{r}_{1}%
)\varphi_{k}^{\ast}(\mathbf{r}_{2})\varphi_{n}(\mathbf{r}_{3})\varphi
_{n}^{\ast}(\mathbf{r}_{4})\varphi_{m}(\mathbf{r}_{4})\varphi_{m}^{\ast
}(\mathbf{r}_{3})}{\omega-\varepsilon_{k}+\varepsilon_{n}-\varepsilon
_{m}+i\delta}\right\}  . \label{Sigmad}%
\end{align}

Comparison between $\Sigma^{d}$ and $\Sigma^{x}$ reveals that the
self-screening terms $n=k$ in $\Sigma^{d}$ for an occupied $\varphi_{k}$,%

\begin{equation}
\frac{\varphi_{k}(\mathbf{r}_{1})\varphi_{k}^{\ast}(\mathbf{r}_{2})\varphi
_{m}(\mathbf{r}_{3})\varphi_{m}^{\ast}(\mathbf{r}_{4})\varphi_{k}%
(\mathbf{r}_{4})\varphi_{k}^{\ast}(\mathbf{r}_{3})}{\omega-\varepsilon
_{k}+\varepsilon_{m}-\varepsilon_{k}-i\delta},
\end{equation}
where $\varphi_{m}$ is unoccupied, are cancelled by the corresponding terms in
$\Sigma^{x}$. Similarly for the case when $\varphi_{k}$ is unoccupied. Thus we
see that by removing the self-screening terms from the direct self-energy we
effectively include the exchange self-energy.

\section{The random-phase approximation with self-polarization correction}

In the previous section we have developed a scheme for removing the
self-screening in the random-phase approximation (RPA) \cite{Pines} in
relation to the $GW$ approximation. When considering the propagation of an
electron or a hole that is screened by the surronding electrons, the electron
or hole in question should not participate in the screening process. Here, we
apply an analogous idea to the case where the perturbation is not due to an
electron or a hole but to a dipole or an electron-hole excitation.

In the RPA the polarization is given by%

\begin{align}
P(\mathbf{r,r}^{\prime};\omega)  &  =\sum_{\alpha}\left\{  \frac{d_{\alpha
}(\mathbf{r})d_{\alpha}^{\ast}(\mathbf{r}^{\prime})}{\omega-\Delta_{\alpha}%
}-\frac{d_{\alpha}(\mathbf{r}^{\prime})d_{\alpha}^{\ast}(\mathbf{r})}%
{\omega+\Delta_{\alpha}}\right\} \nonumber\\
&  =\sum_{\alpha}p_{\alpha}(\mathbf{r,r}^{\prime};\omega),
\end{align}

\begin{align}
d_{\alpha}(\mathbf{r})  &  =\varphi_{m}(\mathbf{r})\varphi_{n}^{\ast
}(\mathbf{r}),\nonumber\\
\Delta_{\alpha}  &  =\varepsilon_{m}-\varepsilon_{n}-i\delta,\ \ \varepsilon
_{m}>\mu,\ \varepsilon_{n}\leq\mu.
\end{align}
The index $\alpha$ includes the spin. The response function is given by%

\begin{align}
R  &  =[1-Pv]^{-1}P\nonumber\\
&  =P+PvP+PvPvP+\cdot\cdot\cdot. \label{RPA}%
\end{align}
We can think of $[1-Pv]^{-1}=\epsilon^{-1}$ as a screening factor that screens
the bare polarization $P$ which consists of electron-hole excitations
$\left\{  p_{\alpha}\right\}  $. We observe that a given electron-hole
excitation $p_{\alpha}$ generates via the Coulomb interaction screening
polarizations that include itself because $P$ contains $p_{\alpha}$. To
eliminate this self-polarization we therefore calculate the self-polarization
corrected response function as follows:%

\begin{equation}
R=\sum_{\alpha}[1-P_{\alpha}v]^{-1}p_{\alpha},
\end{equation}
where%

\begin{equation}
P_{a}=P-p_{\alpha}.
\end{equation}
Physically this means that a particular polarization $p_{\alpha}$ should not
participate again in the screening process so that it should be subtracted out
from $P$. To distinguish it from self-screening, we have referred to this type
of process as "self-polarization" although in essence it is also a
self-screening process.

Analogous to the self-screening correction described before, the
self-polarization correction may be regarded as an appproximate way of
including the exchange diagrams. Consider the first order direct and exchange
terms. The direct term is given by%

\begin{align}
&  P_{d}(x_{1},x_{2})=-\int d^{4}x_{3}d^{4}x_{4}G(x_{3},x_{1})G(x_{1}%
,x_{3}\mathbf{)}\nonumber\\
&  \ \ \ \ \ \ \ \ \ \ \ \ \ \ \ \ \ \ \ \ \ \ \ \ \ \times v(3-4)G(x_{4}%
,x_{2})G(x_{2},x_{4}).
\end{align}
For the exchange term we have%

\begin{align}
&  P_{x}(x_{1},x_{2})=\int d^{4}x_{3}d^{4}x_{4}G(x_{4},x_{1})G(x_{2}%
,x_{4}\mathbf{)}\nonumber\\
&  \ \ \ \ \ \ \ \ \ \ \ \ \ \ \ \ \ \ \ \ \ \ \ \ \ \times G(x_{3}%
,x_{2})G(x_{1},x_{3})v(3-4).
\end{align}
Writing the Green functions in Fourier representation yields%

\begin{align}
&  P_{d}(\mathbf{r}_{1},\mathbf{r}_{2};\omega)\nonumber\\
&  =-\int d^{3}r_{3}d^{3}r_{4}\int\frac{d\omega_{1}}{2\pi}G(\mathbf{r}%
_{3},\mathbf{r}_{1};\omega_{1})G(\mathbf{r}_{1},\mathbf{r}_{3};\omega
_{1}+\omega)\nonumber\\
&  \times\int\frac{d\omega_{3}}{2\pi}G(\mathbf{r}_{2},\mathbf{r}_{4}%
;\omega_{3})G(\mathbf{r}_{4},\mathbf{r}_{2};\omega_{3}+\omega)v(\mathbf{r}%
_{3}-\mathbf{r}_{4})
\end{align}
and%

\begin{align}
&  P_{x}(\mathbf{r}_{1},\mathbf{r}_{2};\omega)\nonumber\\
&  =\int d^{3}r_{3}d^{3}r_{4}\int\frac{d\omega_{1}}{2\pi}G(\mathbf{r}%
_{4},\mathbf{r}_{1};\omega_{1})G(\mathbf{r}_{1},\mathbf{r}_{3};\omega
_{1}+\omega)\nonumber\\
&  \times\int\frac{d\omega_{3}}{2\pi}G(\mathbf{r}_{2},\mathbf{r}_{4}%
;\omega_{3})G(\mathbf{r}_{3},\mathbf{r}_{2};\omega_{3}+\omega)v(\mathbf{r}%
_{3}-\mathbf{r}_{4}).
\end{align}
Using a non-interacting Green function of a given spin yields, using the
convention that repeated indices are summed and $n,n^{\prime}$ refer to the
occupied orbitals whereas $m,m^{\prime}$ to the unoccupied orbitals,%

\begin{align}
P_{d}(\mathbf{r}_{1},\mathbf{r}_{2};\omega)  &  =\frac{\varphi_{n}^{\ast
}(\mathbf{r}_{1})\varphi_{m}(\mathbf{r}_{1})v_{nm,n^{\prime}m^{\prime}}%
\varphi_{n^{\prime}}(\mathbf{r}_{2})\varphi_{m^{\prime}}^{\ast}(\mathbf{r}%
_{2})}{(\omega-\varepsilon_{m}+\varepsilon_{n}+i\delta)(\omega-\varepsilon
_{m^{\prime}}+\varepsilon_{n^{\prime}}+i\delta)}\nonumber\\
&  -\frac{\varphi_{n}^{\ast}(\mathbf{r}_{1})\varphi_{m}(\mathbf{r}%
_{1})v_{nm,m^{\prime}n^{\prime}}\varphi_{m^{\prime}}(\mathbf{r}_{2}%
)\varphi_{n^{\prime}}^{\ast}(\mathbf{r}_{2})}{(\omega-\varepsilon
_{m}+\varepsilon_{n}+i\delta)(\omega+\varepsilon_{m^{\prime}}-\varepsilon
_{n^{\prime}}-i\delta)}\nonumber\\
&  -\frac{\varphi_{m}^{\ast}(\mathbf{r}_{1})\varphi_{n}(\mathbf{r}%
_{1})v_{mn,n^{\prime}m^{\prime}}\varphi_{n^{\prime}}(\mathbf{r}_{2}%
)\varphi_{m^{\prime}}^{\ast}(\mathbf{r}_{2})}{(\omega+\varepsilon
_{m}-\varepsilon_{n}-i\delta)(\omega-\varepsilon_{m^{\prime}}+\varepsilon
_{n^{\prime}}+i\delta)}\nonumber\\
&  +\frac{\varphi_{m}^{\ast}(\mathbf{r}_{1})\varphi_{n}(\mathbf{r}%
_{1})v_{mn,m^{\prime}n^{\prime}}\varphi_{m^{\prime}}(\mathbf{r}_{2}%
)\varphi_{n^{\prime}}^{\ast}(\mathbf{r}_{2})}{(\omega+\varepsilon
_{m}-\varepsilon_{n}-i\delta)(\omega+\varepsilon_{m^{\prime}}-\varepsilon
_{n^{\prime}}-i\delta)}. \label{Pd}%
\end{align}
For the exchange term we obtain for a given spin%

\begin{align}
P_{x}(\mathbf{r}_{1},\mathbf{r}_{2};\omega)  &  =-\frac{\varphi_{n}^{\ast
}(\mathbf{r}_{1})\varphi_{m}(\mathbf{r}_{1})v_{nn^{\prime},mm^{\prime}}%
\varphi_{n^{\prime}}(\mathbf{r}_{2})\varphi_{m^{\prime}}^{\ast}(\mathbf{r}%
_{2})}{(\omega-\varepsilon_{m}+\varepsilon_{n}+i\delta)(\omega-\varepsilon
_{m^{\prime}}+\varepsilon_{n^{\prime}}+i\delta)}\nonumber\\
&  +\frac{\varphi_{n}^{\ast}(\mathbf{r}_{1})\varphi_{m}(\mathbf{r}%
_{1})v_{nm^{\prime},mn^{\prime}}\varphi_{m^{\prime}}(\mathbf{r}_{2}%
)\varphi_{n^{\prime}}^{\ast}(\mathbf{r}_{2})}{(\omega-\varepsilon
_{m}+\varepsilon_{n}+i\delta)(\omega+\varepsilon_{m^{\prime}}-\varepsilon
_{n^{\prime}}-i\delta)}\nonumber\\
&  +\frac{\varphi_{m}^{\ast}(\mathbf{r}_{1})\varphi_{n}(\mathbf{r}%
_{1})v_{mn^{\prime},nm^{\prime}}\varphi_{n^{\prime}}(\mathbf{r}_{2}%
)\varphi_{m^{\prime}}^{\ast}(\mathbf{r}_{2})}{(\omega+\varepsilon
_{m}-\varepsilon_{n}-i\delta)(\omega-\varepsilon_{m^{\prime}}+\varepsilon
_{n^{\prime}}+i\delta)}\nonumber\\
&  -\frac{\varphi_{m}^{\ast}(\mathbf{r}_{1})\varphi_{n}(\mathbf{r}%
_{1})v_{mm^{\prime},nn^{\prime}}\varphi_{m^{\prime}}(\mathbf{r}_{2}%
)\varphi_{n^{\prime}}^{\ast}(\mathbf{r}_{2})}{(\omega+\varepsilon
_{m}-\varepsilon_{n}-i\delta)(\omega+\varepsilon_{m^{\prime}}-\varepsilon
_{n^{\prime}}-i\delta)}, \label{Px}%
\end{align}
where%

\begin{equation}
v_{ij,kl}=\int d^{3}rd^{3}r^{\prime}\varphi_{i}(\mathbf{r})\varphi_{j}^{\ast
}(\mathbf{r})v(\mathbf{r-r}^{\prime})\varphi_{k}^{\ast}(\mathbf{r}^{\prime
})\varphi_{l}(\mathbf{r}^{\prime}).
\end{equation}
The two self-polarization terms, corresponding to $n=n^{\prime}$ and
$m=m^{\prime}$ in the second and third terms of (\ref{Pd}), are cancelled by
the corresponding terms in $P_{x}$.

\section{Application to a model hydrogen dimer}

\subsection{The HOMO-LUMO gap in the conventional GWA}

Consider a model hydrogen molecule with one orbital centered on each atom. The
two orbitals centered on different hydrogen atoms, $\varphi_{1}$ and
$\varphi_{2}$, are normalized but not in general orthogonal: $\left\langle
\varphi_{1}|\varphi_{2}\right\rangle \neq0$. The one-particle eigenfunctions
are the bonding and anti-bonding states:%

\begin{equation}
\psi_{B}=\frac{1}{\sqrt{2}}[\varphi_{1}+\varphi_{2}],
\end{equation}

\begin{equation}
\psi_{A}=\frac{1}{\sqrt{2}}[\varphi_{1}-\varphi_{2}],
\end{equation}
with eigenenergies respectively $\varepsilon_{B}$ and $\varepsilon_{A}$. The
indices $A$ and $B$ include the spin function $\alpha$ and $\beta$. These two
eigenfunctions are orthonormal. We may assume that $\phi_{1}$ and $\phi_{2}$
are real. The two electrons occupy the bonding state with up and down spin.
The non-interacting Green function (the up and down spin Green functions are
identical) is given by%

\begin{equation}
G^{0}(\mathbf{r,r}^{\prime};\omega)=\frac{\psi_{B}(\mathbf{r})\psi
_{B}(\mathbf{r}^{\prime})}{\omega-\varepsilon_{B}-i\delta}+\frac{\psi
_{A}(\mathbf{r})\psi_{A}(\mathbf{r}^{\prime})}{\omega-\varepsilon_{A}+i\delta
},
\end{equation}
where the one-particle Hamiltonian is taken to be the Hartree one. The
HOMO-LUMO gap in the Hartree approximation is%

\begin{equation}
\Delta^{\text{H}}=\varepsilon_{A}-\varepsilon_{B}=2t,
\end{equation}
where the hopping integral is given by%

\begin{equation}
t=-\left\langle \varphi_{1}|-\frac{1}{2}\nabla^{2}+v_{\text{ext}}+V_{\text{H}%
}|\varphi_{2}\right\rangle .
\end{equation}
The onsite and intersite Coulomb interactions are respectively{
\begin{align}
U_{0}  &  =\left\langle \varphi_{1}^{2}|v|\varphi_{1}^{2}\right\rangle
=\left\langle \varphi_{2}^{2}|v|\varphi_{2}^{2}\right\rangle ,\\
U_{1}  &  =\left\langle \varphi_{1}^{2}|v|\varphi_{2}^{2}\right\rangle .
\end{align}
}$\left\langle \varphi_{1}\varphi_{2}|v|\varphi_{1}\varphi_{2}\right\rangle $
and $\left\langle \varphi_{1}^{2}|v|\varphi_{1}\varphi_{2}\right\rangle $ are
neglected since they are much smaller compared with $U_{0}$ and $U_{1}$. { }

First, let us calculate the exchange contribution:%

\begin{equation}
\Sigma^{\text{x}}(\mathbf{r,r}^{\prime})=-v(\mathbf{r-r}^{\prime})\psi
_{B}(\mathbf{r})\psi_{B}(\mathbf{r}^{\prime}).
\end{equation}
The matrix elements in the bonding and anti-bonding states are%

\begin{align}
\left\langle \psi_{B}|\Sigma^{\text{x}}|\psi_{B}\right\rangle  &  =-\frac
{1}{2}(U_{0}+U_{1}),\\
\left\langle \psi_{A}|\Sigma^{\text{x}}|\psi_{A}\right\rangle  &  =-\frac
{1}{2}(U_{0}-U_{1}).
\end{align}
The HOMO-LUMO gap in the HFA is therefore%

\begin{equation}
\Delta^{\text{HF}}=2t+U_{1}. \label{HFgap}%
\end{equation}

We now proceed to calculate the correlation part of the self-energy. The
polarization function can be written in the form%

\begin{equation}
P^{0}(\mathbf{r,r}^{\prime};\omega)=\psi_{B}(\mathbf{r})\psi_{A}%
(\mathbf{r})P^{0}(\omega)\psi_{B}(\mathbf{r}^{\prime})\psi_{A}(\mathbf{r}%
^{\prime}),
\end{equation}
where%

\begin{equation}
P^{0}(\omega)=2\left\{  \frac{1}{\omega-\Delta\varepsilon+i\delta}-\frac
{1}{\omega+\Delta\varepsilon-i\delta}\right\}  \label{P0}%
\end{equation}
with%

\begin{equation}
\Delta\varepsilon=\varepsilon_{A}-\varepsilon_{B}.
\end{equation}
The factor of $2$ in (\ref{P0}) is due to the sum over spin. Using the RPA
equation in (\ref{RPA}) and solving it by iteration, it is straightforward to
see that each term in the iterative solution can be written in the same form
as $P^{0}$ so that the response function can also be written as%

\begin{equation}
R(\mathbf{r,r}^{\prime};\omega)=\psi_{B}(\mathbf{r})\psi_{A}(\mathbf{r}%
)R(\omega)\psi_{B}(\mathbf{r}^{\prime})\psi_{A}(\mathbf{r}^{\prime}).
\end{equation}
$R(\omega)$ can be calculated algebraically and it is given by%

\begin{equation}
R(\omega)=\frac{2r}{\omega-\Delta E+i\delta}-\frac{2r}{\omega+\Delta
E-i\delta},
\end{equation}
where%

\begin{equation}
\Delta E=\sqrt{(\Delta\varepsilon)^{2}+4v_{BA,BA}\Delta\varepsilon,}%
\end{equation}

\begin{equation}
v_{ab,cd}=\int d^{3}rd^{3}r^{\prime}\psi_{a}(\mathbf{r})\psi_{b}%
(\mathbf{r})v(\mathbf{r-r}^{\prime})\psi_{c}(\mathbf{r}^{\prime})\psi
_{d}(\mathbf{r}^{\prime}),
\end{equation}

\begin{align}
r  &  =\frac{\Delta\varepsilon}{\Delta E}<1,\nonumber\\
&  =\frac{1}{\sqrt{(1+\frac{2(U_{0}-U_{1})}{\Delta\varepsilon}}}\nonumber\\
&  \approx1-\frac{U_{0}-U_{1}}{2t}.
\end{align}

Using%

\begin{equation}
W^{c}(\mathbf{r}^{\prime}\mathbf{,r};\omega)=\int d^{3}r_{1}d^{3}%
r_{2}v(\mathbf{r}^{\prime}\mathbf{-r}_{1})R(\mathbf{r}_{1},\mathbf{r}%
_{2};\omega)v(\mathbf{r}_{2}-\mathbf{r}),
\end{equation}
the correlation part of the self-energy $\Sigma^{c}$ can be calculated
analytically to yield%

\begin{align}
\Sigma_{GW}^{c}(\mathbf{r,r}^{\prime};\omega)  &  =i\int\frac{d\omega^{\prime
}}{2\pi}G^{0}(\mathbf{r,r}^{\prime};\omega+\omega^{\prime})W^{c}%
(\mathbf{r}^{\prime},\mathbf{r};\omega^{\prime})\nonumber\\
&  =\frac{\lambda_{1}(\mathbf{r,r}^{\prime})}{\omega+\Delta E-\varepsilon
_{B}-i\delta}+\frac{\lambda_{2}(\mathbf{r,r}^{\prime})}{\omega-\Delta
E-\varepsilon_{A}+i\delta},
\end{align}
where%

\begin{align}
&  \lambda_{1}(\mathbf{r,r}^{\prime})=2r\psi_{B}(\mathbf{r)}\psi
_{B}(\mathbf{r}^{\prime})\int d^{3}r_{1}d^{3}r_{2}v(\mathbf{r-r}%
_{1})\nonumber\\
&  \times\psi_{B}(\mathbf{r}_{1})\psi_{A}(\mathbf{r}_{1})\psi_{A}%
(\mathbf{r}_{2})\psi_{B}(\mathbf{r}_{2})v(\mathbf{r}_{2}-\mathbf{r}^{\prime}),
\label{lambda1}%
\end{align}

\begin{align}
&  \lambda_{2}(\mathbf{r,r}^{\prime})=2r\psi_{A}(\mathbf{r)}\psi
_{A}(\mathbf{r}^{\prime})\int d^{3}r_{1}d^{3}r_{2}v(\mathbf{r-r}%
_{1})\nonumber\\
&  \times\psi_{B}(\mathbf{r}_{1})\psi_{A}(\mathbf{r}_{1})\psi_{A}%
(\mathbf{r}_{2})\psi_{B}(\mathbf{r}_{2})v(\mathbf{r}_{2}-\mathbf{r}^{\prime}).
\label{lambda2}%
\end{align}

\begin{align}
&  \left\langle \psi_{B}|\Sigma_{GW}^{c}(\omega)|\psi_{B}\right\rangle
\nonumber\\
&  =\frac{2rv_{BB,AB}^{2}}{\omega+\Delta E-\varepsilon_{B}-i\delta}%
+\frac{2rv_{AB,AB}^{2}}{\omega-\Delta E-\varepsilon_{A}+i\delta}\nonumber\\
&  =\frac{1}{2}\frac{r(U_{0}-U_{1})^{2}}{\omega-\Delta E-\varepsilon
_{A}+i\delta}, \label{GWbond}%
\end{align}

\begin{align}
&  \left\langle \psi_{A}|\Sigma_{GW}^{c}(\omega)|\psi_{A}\right\rangle
\nonumber\\
&  =\frac{2rv_{AB,AB}^{2}}{\omega+\Delta E-\varepsilon_{B}-i\delta}%
+\frac{2rv_{AA,AB}^{2}}{\omega-\Delta E-\varepsilon_{A}+i\delta}\nonumber\\
&  =\frac{1}{2}\frac{r(U_{0}-U_{1})^{2}}{\omega+\Delta E-\varepsilon
_{B}-i\delta}. \label{GWantibond}%
\end{align}
Adding to the Hartree-Fock gap in (\ref{HFgap}) the HOMO-LUMO gap in the GWA
is therefore%

\begin{equation}
\Delta^{GW}=2t+U_{1}+\frac{r(U_{0}-U_{1})^{2}}{\Delta\varepsilon+\Delta E}.
\label{GW_gap}%
\end{equation}
It is interesting to note that correlation effects increase the Hartree-Fock
gap, counter to the usual expectation.

\subsection{The HOMO-LUMO gap in the GWA with self-screening correction}

Let us now apply our $GW$ with self-screening correction scheme. For an
electron of a given spin in the bonding or anti-bonding state the screening is
provided by the other electron with opposite spin, as it should. Thus, the
polarization is half of the polarization without self-screening correction.
The calculation proceeds as in the previous section and we obtain%

\begin{equation}
R(\omega)=\frac{r}{\omega-\Delta E+i\delta}-\frac{r}{\omega+\Delta E-i\delta},
\end{equation}
where%

\begin{equation}
\Delta E=\sqrt{(\Delta\varepsilon)^{2}+2v_{BA,BA}\Delta\varepsilon
},\ \ r=\frac{\Delta\varepsilon}{\Delta E}.
\end{equation}
The correlation part of the self-energy with self-screening correction is%
\begin{align}
\Sigma_{GW\text{-SS}}^{c}(\mathbf{r,r}^{\prime};\omega)  &  =\frac{1}{2}%
\frac{\lambda_{1}(\mathbf{r,r}^{\prime})}{\omega+\Delta E-\varepsilon
_{B}-i\delta}\nonumber\\
&  +\frac{1}{2}\frac{\lambda_{2}(\mathbf{r,r}^{\prime})}{\omega-\Delta
E-\varepsilon_{A}+i\delta},
\end{align}
where $\lambda_{1}$ and $\lambda_{2}$ are given in (\ref{lambda1}) and
(\ref{lambda2}). Thus,%

\begin{align}
&  \left\langle \psi_{B}|\Sigma_{GW\text{-SS}}^{c}(\omega)|\psi_{B}%
\right\rangle \nonumber\\
&  =\frac{rv_{BB,AB}^{2}}{\omega+\Delta E-\varepsilon_{B}-i\delta}%
+\frac{rv_{AB,AB}^{2}}{\omega-\Delta E-\varepsilon_{A}+i\delta}\nonumber\\
&  =\frac{1}{4}\frac{r(U_{0}-U_{1})^{2}}{\omega-\Delta E-\varepsilon
_{A}+i\delta}, \label{bonding}%
\end{align}

\begin{align}
&  \left\langle \psi_{A}|\Sigma_{GW\text{-SS}}^{c}(\omega)|\psi_{A}%
\right\rangle \nonumber\\
&  =\frac{rv_{AB,AB}^{2}}{\omega+\Delta E-\varepsilon_{B}-i\delta}%
+\frac{rv_{AA,AB}^{2}}{\omega-\Delta E-\varepsilon_{A}+i\delta}\nonumber\\
&  =\frac{1}{4}\frac{r(U_{0}-U_{1})^{2}}{\omega+\Delta E-\varepsilon
_{B}-i\delta}. \label{antibonding}%
\end{align}
Taking into account the Hartree-Fock gap in (\ref{HFgap}), the
self-screening-corrected $GW$ HOMO-LUMO gap is therefore%

\begin{equation}
\Delta^{GW\text{-SS}}=2t+U_{1}+\frac{r(U_{0}-U_{1})^{2}}{2(\Delta
\varepsilon+\Delta E)},
\end{equation}
where%

\begin{equation}
r=\frac{\Delta\varepsilon}{\Delta E}=\left[  1+\frac{U_{0}-U_{1}}%
{\Delta\varepsilon}\right]  ^{-1/2}.
\end{equation}
It is shown below that this is the same as the exact result up to second order
in $(U_{0}-U_{1})/2t$ in the weak to moderate coupling regime where
$(U_{0}-U_{1})/2t<1$.

\subsection{{Exact solution in atomic basis}}

{We consider configurations with total }$S_{z}=0${. In this case the
Hamiltonian is given by%
\begin{equation}
H=\left(
\begin{array}
[c]{cccc}%
2\varepsilon_{0}+U_{1} & 0 & -t & -t\\
0 & 2\varepsilon_{0}+U_{1} & t & t\\
-t & t & 2\varepsilon_{0}+U_{0} & 0\\
-t & t & 0 & 2\varepsilon_{0}+U_{0}%
\end{array}
\right)  ,
\end{equation}
which can be solved analytically. Since $\varepsilon_{0}$ appears only in the
diagonal element, we may set it to zero. Choosing }$\varepsilon_{0}=0${, the
ground-state energy is given by
\begin{equation}
E_{0}(N)=\frac{1}{2}(U_{0}+U_{1})-\frac{1}{2}\sqrt{(U_{0}-U_{1})^{2}+16t^{2}}.
\end{equation}
}

To calculate the bonding-antibonding or HOMO-LUMO gap, we need to consider the
$N\pm1$ problems. For the one- and three-electron problem there are only two
configurations{. }The eigenvalues are%

\begin{align}
E_{1,2}(N+1)  &  =3\varepsilon_{0}+U_{0}+2U_{1}\pm(-t),\\
E_{1,2}(N-1)  &  =\varepsilon_{0}\pm t.
\end{align}

{The exact HOMO-LUMO gap with }$t>0$ {is}%

\begin{align}
\Delta^{\text{exact}}  &  =E_{1}(N+1)-2E_{0}(N)+E_{1}(N-1)\nonumber\\
&  =-2t+U_{1}+\sqrt{(U_{0}-U_{1})^{2}+16t^{2}}.
\end{align}
It approaches $2t$ as $U_{0,1}\rightarrow0$, as it should. In the weak or
moderate coupling regime where $(U_{0}-U_{1})/2t<1$ and the gap is given by%

\begin{align}
\Delta^{\text{exact}}  &  =-2t+U_{1}+\sqrt{(U_{0}-U_{1})^{2}+16t^{2}%
}\nonumber\\
&  \approx2t+U_{1}+\frac{t}{2}\left(  \frac{U_{0}-U_{1}}{2t}\right)  ^{2}.
\end{align}
This is the same as the gap in the $GW$ scheme with self-screening correction
up to order $[(U_{0}-U_{1})/2t]^{2}$:%

\begin{align}
\Delta^{GW\text{-SS}}  &  =2t+U_{1}+\frac{r(U_{0}-U_{1})^{2}}{2(\Delta
\varepsilon+\Delta E)}\nonumber\\
&  \approx2t+U_{1}+\frac{t}{2}\left(  \frac{U_{0}-U_{1}}{2t}\right)  ^{2}.
\end{align}

\section{Conclusion}

We have proposed a scheme for taking into account self-screening correction
within the GWA. The scheme introduces orbital and spin dependent screened
interaction. While this is not necessary in theory, the introduction of
orbital and spin dependence $W$ within the GWA captures the essential physics
better and improves the self-energy without resorting to complicated vertex
corrections. This is analogous to the introduction of the spin variable in the
spin density functional theory. The scheme is justified theoretically by
showing that the self-screening terms are indeed cancelled when exchange
diagrams beyond the GWA are considered. When applied to a model hydrogen
dimer, the scheme reproduces the exact result in the weak to moderate coupling
regime. Work is now under way to apply the scheme to real systems.

\section{Acknowledgment}

FA would like to thank Olle Gunnarsson from fruitfull discussions. FA
acknowledges support from G-COE program of MEXT (G-03) Japan.

\end{document}